\documentclass[letterpaper]{article} 
\usepackage{aaai2026}  
\usepackage{times}  
\usepackage{helvet}  
\usepackage{courier}  
\usepackage[hyphens]{url}  
\usepackage{graphicx} 
\usepackage{natbib}  
\usepackage{caption} 
\usepackage{algorithm}

\usepackage{algorithmic}
\usepackage{amsmath}
\usepackage{subfigure}
\usepackage{booktabs} 

\usepackage{newfloat}
\usepackage{listings}
\DeclareCaptionStyle{ruled}{labelfont=normalfont,labelsep=colon,strut=off} 
\floatstyle{ruled}
\newfloat{listing}{tb}{lst}{}
\floatname{listing}{Listing}
\nocopyright

\title{Camel: Energy-Aware LLM Inference on Resource-Constrained Devices}
\author{
    Hao Xu, Long Peng\thanks{Corresponding author. penglong@nudt.edu.cn}, Shezheng Song, Xiaodong Liu, Ma Jun, Shasha Li, Jie Yu, Xiaoguang Mao
}
\affiliations{
    College of Computer Science and Technology, National University of Defense Technology\\

}

\usepackage{bibentry}

\begin{document}

\maketitle

\begin{abstract}
Most Large Language Models (LLMs) are currently deployed in the cloud, with users relying on internet connectivity for access. However, this paradigm faces challenges such as network latency, privacy concerns, and bandwidth limits. Thus, deploying LLMs on edge devices has become an important research focus. In edge inference, request latency is critical as high latency can impair real-time tasks. At the same time, edge devices usually have limited battery capacity, making energy consumption another major concern. Balancing energy consumption and inference latency is essential.
To address this, we propose an LLM inference energy management framework that optimizes GPU frequency and batch size to balance latency and energy consumption. By effectively managing the exploration-exploitation dilemma in configuration search, the framework finds the optimal settings. The framework was implemented on the NVIDIA Jetson AGX Orin platform, and a series of experimental validations were conducted. Results demonstrate that, compared to the default configuration, our framework reduces energy delay product (EDP) by 12.4\%-29.9\%, achieving a better balance between energy consumption and latency.
\end{abstract}


\section{Introduction}
Large language models (LLMs) have achieved remarkable advancements in areas such as natural language processing and image processing, paving the way for the realization of general artificial intelligence. Nevertheless, inference of LLMs requires enormous energy consumption. For instance, ChatGPT \cite{chatgpt} consumes over 500,000 kWh of electricity daily to handle around 200 million requests, which is more than 17,000 times the daily electricity use of an average U.S. household (29 kWh) \cite{smith2021rise}.

Most LLMs run on cloud servers accessed via networks, but this approach faces issues like network latency, privacy, and security concerns. To address these, deploying LLMs on \textbf{edge devices} for local inference is gaining attention. Edge computing moves resources closer to users, reducing network latency and bandwidth demands, lessening reliance on cloud servers. Latency is a crucial factor to consider for edge LLM inference. For systems that require real-time responses, high latency can directly hinder task execution and may even lead to a significant decrease in system performance. However, edge devices have limited power, making energy consumption a critical issue that cannot be ignored.



\textbf{This paper investigates the tradeoff between energy consumption and inference latency for LLM inference on edge devices.}  In this context, requests are processed sequentially in batches, where batch size (the number of requests handled simultaneously) is crucial. Larger batches improve inference parallelism, but since requests are handled sequentially, a larger batch size leads to longer waiting times before processing. Conversely, smaller batches reduce waiting time for individual requests, but limit parallelism. GPU frequency also affects performance: higher frequency lowers latency but raises energy consumption, while lower frequency reduces energy consumption at the cost of higher latency. Finding the optimal balance between latency and energy consumption is a key challenge explored in this study.

\begin{figure}[htb]
\centering
    \includegraphics[width=0.7\linewidth]{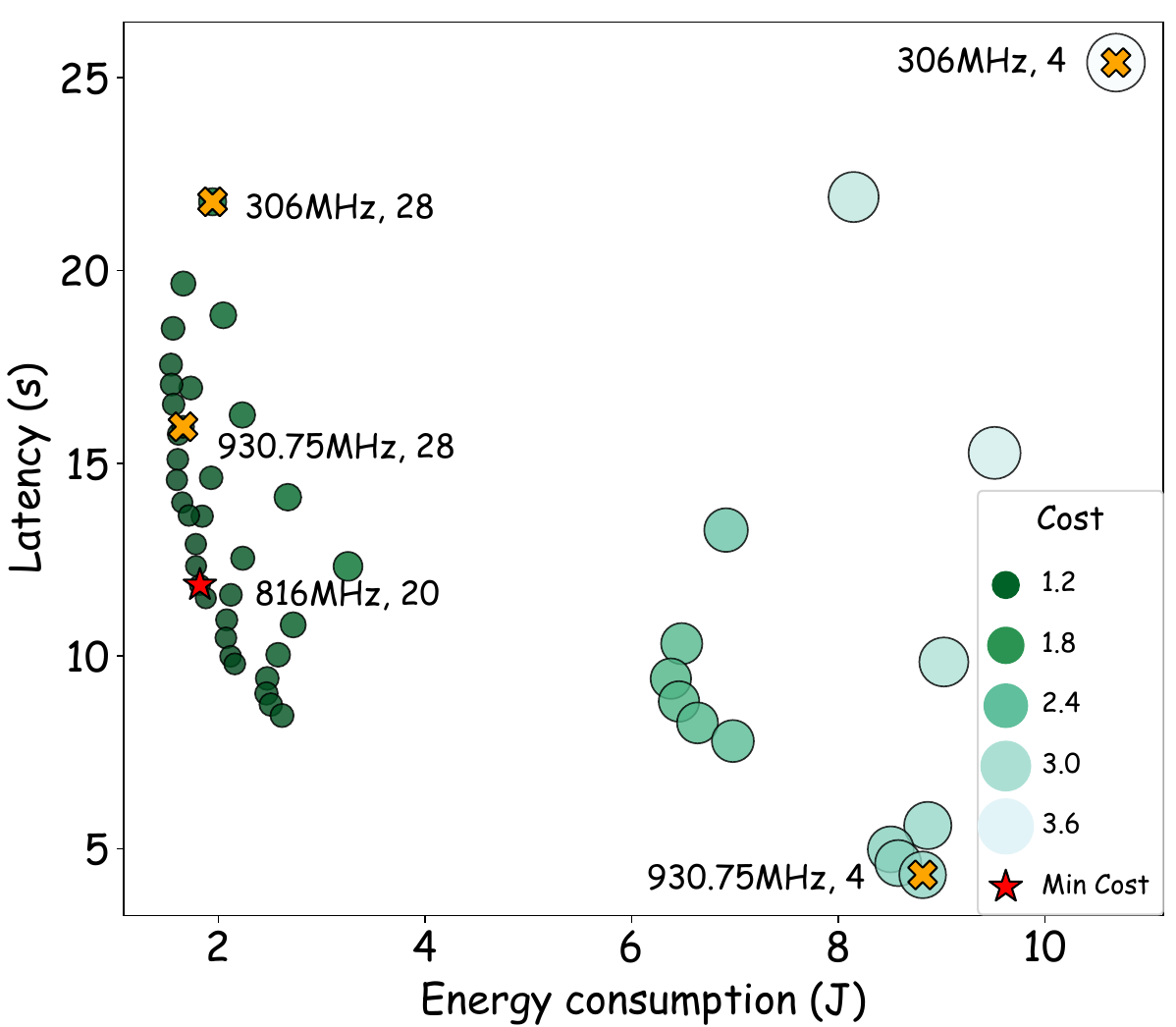}
    \caption{The average energy consumption, latency, and cost per request under different GPU frequencies and batch sizes. The red star marks the combination with the lowest cost, while the yellow ``X'' marker indicates configurations with the maximum and minimum values across the four GPU frequencies and batch sizes.}
    \label{fig:observation}
\end{figure}

To better understand the tradeoff between latency and energy consumption during LLM inference on edge devices, we conducted experiments on the Jetson AGX Orin with the Llama3.2-1B model. Request arrivals were fixed at 1-second intervals, while GPU frequency varied over seven discrete levels from 306 MHz to 930.75 MHz. Batch size was also adjusted over seven levels, from 4 to 28, resulting in 49 unique configurations. Fig. \ref{fig:observation} shows the average latency (time from request arrival to completion) and energy consumption per request for each configuration. To evaluate their tradeoff, we defined a metric cost as the weighted sum of energy consumption and latency, with equal weights (0.5 each). We normalized the cost metric by following the work \cite{Rachuri2024191}, setting the cost corresponding to the highest GPU frequency and the highest batch size to 1. Latency and energy for each configuration were divided by these baseline values to compute the cost value. In Fig. \ref{fig:observation}, the size of each circle represents the corresponding cost.

Besides the configuration with minimum cost (816 MHz, 20), four other representative configurations are labeled in Fig. \ref{fig:observation}:
(930.75 MHz, 4), (930.75 MHz, 28), (306 MHz, 4), and (306 MHz, 28).
Results reveal that the optimal GPU frequency and batch size for LLM inference on edge devices do not necessarily lie at the extremes, highlighting the need for precise algorithmic optimization to find the best configuration.

To effectively address the aforementioned challenges, this paper proposes an energy management framework for LLM inference, named Camel. Our goal is to achieve the optimal balance between latency and energy consumption by selecting the optimal GPU frequency (system-level) and batch size (application-level). To this end, we model the problem as a Multi-armed Bandit (MAB) problem, where the GPU frequency and batch size during LLM inference are treated as decision options, and the inference time and latency corresponding to each decision constitute its associated cost. Inspired by Thompson Sampling \cite{ts}, we establish two different Gaussian distributions for each decision: 1) the cost resulting from the decision follows a Gaussian distribution with a mean of $\theta$; 2) the mean $\theta$ itself follows another Gaussian distribution. By repeatedly sampling from different decision options and dynamically updating the Gaussian distribution of the mean $\theta$ based on cost, our framework effectively balances exploration and exploitation, ultimately converging to the optimal decision.

We have implemented the Camel framework on the NVIDIA Jetson AGX Orin platform, selecting Llama.cpp \cite{llmacpp} as the inference framework. Experimental results show that compared to the default GPU frequency and batch size configuration, Camel reduces energy delay product (EDP) \cite{EDP1} by 12.4\%-29.9\% for the Llama3.2-1B and Qwen2.5-3B models, achieving a better balance between energy consumption and latency.

The contributions of this paper are as follows:
\begin{itemize}
    \item To the best of our knowledge, our work is the first to explore the tradeoff between latency and energy consumption in LLM batching inference under resource-constrained conditions.
    \item We model the problem of selecting GPU frequency and batch size for LLM inference optimization as an MAB problem, and propose an optimization framework that successfully achieves the optimal balance between latency and energy consumption.
    \item Our method has been implemented on the NVIDIA Jetson AGX Orin development board and validated on mainstream LLMs. It significantly reduces EDP (12.4\%-29.9\%) compared to the default configuration.
\end{itemize}

\section{Related Work}
\subsection{Deep Learning Energy Optimization.}
Many studies focus on the power consumption problem during DNN training, often leveraging DVFS technology to adjust the device frequency and achieve an optimal balance between power consumption and performance \cite{Wang20222943,Bharadwaj202149,Zou2020559,Tang2019315,Komoda2013349,zeus}. CE-DLA\cite{Kang2022} reduces the energy consumption of clusters by efficiently scheduling deep learning tasks to appropriate GPU computing nodes. Work\cite{Komoda2013349} introduces an effective power-limiting technique that integrates DVFS with task mapping, specifically tailored for CPU-GPU heterogeneous systems. Zeus\cite{zeus}, which shares similarities with our approach, also frames the problem as a MAB problem to select the optimal training configuration. However, unlike these studies, our research focuses on the inference phase for LLMs.

Some other studies have concentrated on optimizing power consumption during DNN inference on edge devices \cite{Dutt202352,Jeong2022,Rachuri2024191}. BatchSizer \cite{Nabavinejad2021819} achieves efficient DNN inference under power constraints by dynamically controlling the input batch size. EENet \cite{Li2023} enhances power efficiency through an early exit strategy, which involves dynamically adjusting the voltage and frequency of the processor by predicting early exit points and calibrating based on inference workload and time constraints. EcoEdgeInfer\cite{Rachuri2024191} is the most similar to our approach, utilizing a gradient-based method to determine the optimal workload and device frequency during DNN inference. However, our work specifically focuses on the LLM scenario.

\subsection{LLM Inference Optimization.}
Model compression encompasses a variety of techniques aimed at enhancing inference efficiency by modifying the data representation of pre-trained models (e.g., quantization \cite{gptq,awq}) or altering their architecture (e.g., pruning \cite{Sparsegpt, Llm-pruner, simplepruner}, distillation \cite{KD1,KD2}, etc.). Speculative sampling \cite{SD1,SD2,SD3} improves the decoding efficiency of autoregressive LLMs by utilizing a smaller model, referred to as a draft model, to predict multiple subsequent tokens in an efficient manner. Computation offloading \cite{flexgen,llmacpp,fastdecode,powerinfer} involves transferring part of the computational workload from the GPU to the CPU when the latter is idle, thereby optimizing resource utilization.
\cite{clone} studies the tradeoff between power consumption and latency during LLM inference, but did not consider the scenario where requests arrive sequentially for service.




\section{Camel Design}

\subsection{Problem Formulation}

We define the average latency  \( L(t) \) for the \(t\)-th batch processing as the duration from the arrival of a request at the server until its complete processing. Similarly, we define the energy consumption \( E(t)\) as the average GPU energy consumed per request throughout this batch processing. Consequently, the problem of minimizing  both \( E(t)\) and \( L(t) \) during the LLM inference is formulated as follows:

\begin{equation}
    \begin{aligned}
    \textbf{(P)}: \min_{f, b} \sum_{t=1}^{T} \{\alpha \cdot E(t)+ (1-\alpha) \cdot {L(t)}\} \\
    \text{s.t.}  \quad f \in \mathcal{F}, \quad b \in \mathcal{B}, \quad \forall t \in [1, T], \quad \alpha \in [0,1]
    \end{aligned}
    \label{eq:objective}
\end{equation}
where \(T\) represents the total number of batch processing iterations, while \(f\) and \(b\) correspond to the GPU frequency and batch size, respectively, both of which are key parameters for adjustment. The parameter \(\alpha\) ranges from 0 to 1, providing flexibility to be adjusted based on user preferences. When \(\alpha=0\), the algorithm primarily focuses on optimizing inference latency, whereas when \(\alpha=1\), the focus shifts to minimizing energy consumption.

We analyze the energy consumption $E$ and latency $L$ as functions of GPU frequency $f$ and batch size $b$.
The total GPU power consumption consists of static and dynamic parts:
\begin{equation}
    \begin{aligned}
    P_{\mathrm{total}}
    = P_{\mathrm{0}} + C\,V(f)^2\,f
    \end{aligned}
    \label{eq:energy_consumption}
\end{equation}
where $P_0$ is static power and dynamic power depends on capacitance \(C\), frequency \(f\), and voltage \(V(f)\).


The batch inference latency for LLMs can be conceptualized as the ratio of the computational load to the GPU frequency \(f\). The computational load consists of two main components: fixed overheads during batch processing \(C_0\), which includes scheduling, initialization, and I/O operations, and the variable computational load per request \(c_p\).

The batch inference latency is:

\begin{equation}
    \begin{aligned}
    t_{\mathrm{batch}}
= \frac{C_0 +b\,c_p}{\mu f}
    \end{aligned}
    \label{eq:batch_t}
\end{equation}
where the \(\mu\) serves only as an empirical fitting parameter without specific physical meaning.

The total energy consumption per batch is:

\begin{equation}
\label{eq:batch_energy}
\begin{aligned}
E_{\mathrm{batch}}
&= P_{total}\cdot t_{\mathrm{batch}}= \left(P_{0} + C\,V(f)^{2}\,f\right)\;\frac{C_0+b\,c_p}{\mu f}
\end{aligned}
\end{equation}

The energy consumed per request is:
\begin{equation}
    \begin{aligned}
E_{\mathrm{request}}
= \frac{E_{\mathrm{batch}}}{b}
= \left(P_{0} + C\,V(f)^{2}\,f\right)\;\frac{C_0+b\,c_p}{b\,\mu f}
    \end{aligned}
    \label{eq:requet_energy}
\end{equation}

Request latency includes queuing wait time \(t_{wait}\) and batch processing time \(t_{batch}\). Requests arrive at the server at a uniform rate of $\lambda$, and the server begins batch processing only after accumulating \(b\) requests. In this case, each request, on average, waits for \((b-1)/2\lambda\) seconds before triggering processing along with the \(b\)-th request. Consequently, the average waiting time is:

\begin{equation}
  t_{\mathrm{wait}} = \frac{b - 1}{2\,\lambda}
  \label{eq:waiting_time}
\end{equation}

Thus, total request latency is:

\begin{equation}
\label{eq:total_latency}
\begin{aligned}
L_{request}
&= t_{wait} + t_{batch} = \frac{b - 1}{2\,\lambda} + \frac{C_0 +b\,c_p}{\mu f}
\end{aligned}
\end{equation}

The objective function in Eq. \ref{eq:objective} can be expressed as:

\begin{equation}
\begin{aligned}
\alpha\,E_{\text{request}} + (1-\alpha)\, L_{\text{request}}  =&\ \alpha(P_{0} +  CV(f)^{2}f)\frac{C_0+bc_p}{b\mu f} \\
&+ (1-\alpha)(\frac{b - 1}{2\lambda} + \frac{C_0 +bc_p}{\mu f})
\end{aligned}
  \label{eq:total_objective}
\end{equation}

This objective is nonlinear in \(f\) and \(b\), making the determination of an optimal GPU frequency and batch size combination a complex challenge subject to the defined constraints.


\subsection{Camel Overview}
Inspired by Thompson Sampling, we propose the Camel architecture to balance energy consumption and latency. The objective function in Eq. \ref{eq:objective} defines the cost for each ``arm", which represents a combination of GPU frequency and batch size. Unlike classical MAB problems that maximize reward, our goal is to minimize this cost.  


We model the potential cost of each arm as a Gaussian distribution \( N(\theta, \sigma_1^2) \),  where \(\theta\) itself follows a Gaussian distribution  \( N(\mu, \sigma_2^2) \). By sampling repeatedly from \( N(\theta, \sigma_1^2) \), the posterior distribution \( N(\mu, \sigma_2^2) \) is updated continuously, refining the decision-making. Camel integrates prior knowledge to balance exploration and exploitation, avoiding local optima and efficiently identifying the optimal GPU frequency and batch size.

\begin{figure*}
    \centering
    \includegraphics[width=\linewidth]{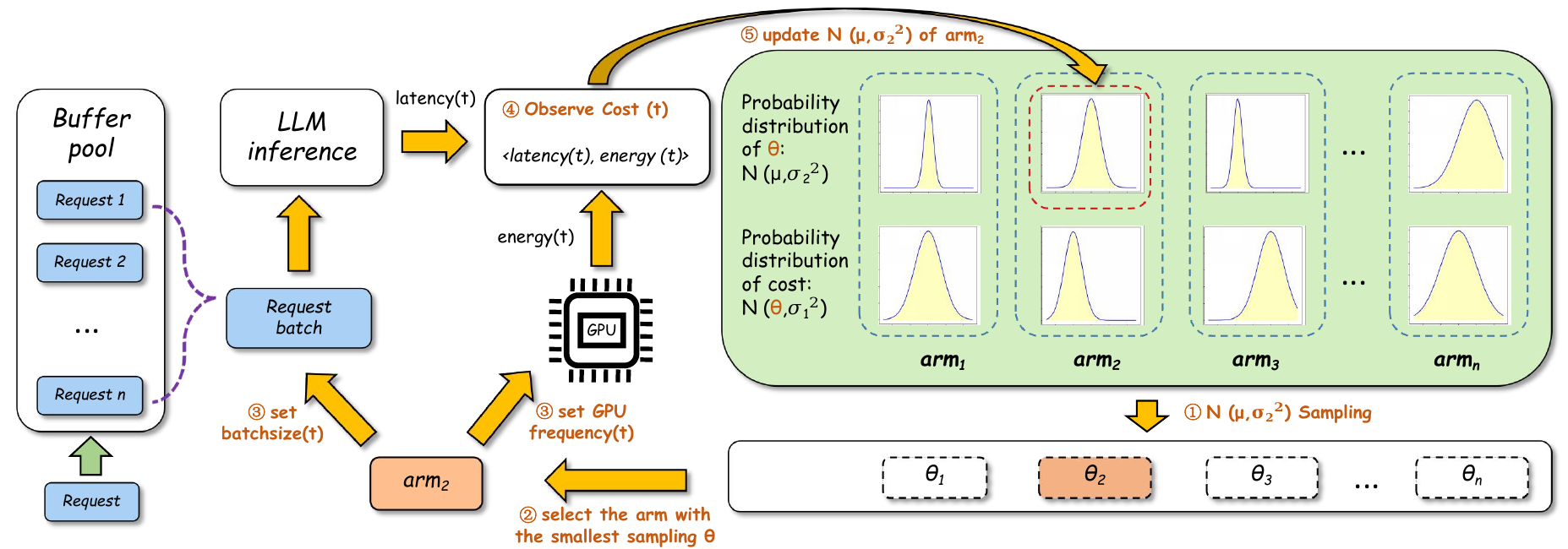}
    \caption{Overview of Camel implementation.}
    \label{fig:overview}
\end{figure*}

As shown in Fig. \ref{fig:overview}, Camel sample $\theta$ for each arm and select the arm with the lowest $\theta$ for the next action (if the sample mean of an arm is smaller, it is reasonable to assume that its cost sample will also be smaller), which includes adjusting GPU frequency and the number of requests to process (batch size). After processing, the observed cost updates the posterior distribution of $\theta$ for the corresponding arm, guiding the next iteration.

\textbf{The core of Camel lies in the continuous post-event update that uses prior information, ensuring decisions that better reflect real-world observations}. The next section details this posterior update process.









\subsection{Update Posterior Distribution}{\label{section:posterior}}

\subsubsection{Bayesian Viewpoint}



Consider \( x = (x_1, x_2, \dots, x_n) \) as a set of samples drawn from \( X \). From the Bayesian perspective, \( X \) depends on the parameter \( \theta \), with the conditional distribution \( p(x | \theta) \). The parameter \( \theta \) is treated as a random variable with a prior distribution \( \pi(\theta) \). The sample \(x\)  is generated by first sampling \( \theta \) from \( \pi(\theta) \), then sampling \( x = (x_1, x_2, \dots, x_n) \) from \( p(x | \theta) \). The likelihood function is:

\begin{equation}
    \begin{aligned}
    p(x | \theta) = \prod_{i=1}^{n} p(x_i | \theta)
    \end{aligned}
    \label{delta_ceta}
\end{equation}


The posterior distribution \( \pi(\theta |x ) \) is proportional to the product of the likelihood function and the prior distribution:

\begin{equation}
    \begin{aligned}
    \pi(\theta | x) \propto p(x | \theta) \pi(\theta)
    \end{aligned}
    \label{delta_ceta}
\end{equation}
The marginal probability \( p(x) \) is obtained by integrating over \( \theta \):

\begin{equation}
    \begin{aligned}
    p(x) = \int p(x | \theta) \pi(\theta) d\theta
    \end{aligned}
    \label{delta_ceta}
\end{equation}

\( p(x) \) is also referred to as the normalization constant, ensuring that the posterior distribution sums to 1. Consequently, the posterior distribution \( \pi(\theta | x) \)  is given by:

\begin{equation}
    \begin{aligned}
    \pi (\theta | x) = \frac{p(x | \theta) \pi(\theta)}{p(x)}
\end{aligned}
    \label{poster_dis}
\end{equation}

\subsubsection{Update Posterior Distribution}

In the process of calculating the posterior distribution, there exists a special scenario in which the prior distribution \( \pi(\theta) \) and the posterior distribution \( \pi(\theta | x) \) belong to the same family of distributions. In this case, \( \pi(\theta) \) is referred to as the conjugate prior distribution for the parameter \( \theta \). 

Assume \( x = \{x_1, x_2, \dots, x_n\} \) is drawn from \( \mathcal{N}(\theta, \sigma_1^2) \) with known variance. The likelihood function is:

\begin{equation}
    \begin{aligned}
    p (x | \theta) =(\frac{1}{\sqrt{2\pi}{\sigma_1}})^n \exp\{{-\frac{1}{2\sigma_1^2}\sum_{i=1}^{n}(x_i-\theta)^2}\}
\end{aligned}
    \label{delta_ceta}
\end{equation}


The prior distribution $\pi (\theta)$ is $\mathcal{N}(\mu, \sigma_2^2)$:

\begin{equation}
    \begin{aligned}
    \pi (\theta) =\frac{1}{\sqrt{2\pi}{\sigma_2}} \exp\{{-\frac{1}{2\sigma_2^2}(\theta-\mu)^2}\}
\end{aligned}
    \label{delta_ceta}
\end{equation}


The joint distribution of  \(x\) and  \(\theta\) is:

\begin{equation}
    \begin{aligned}
    p(x | \theta) \pi(\theta) &= \alpha_1\exp \{{-\frac{1}{2}}(A\theta^2-2B\theta+ C)\}
\end{aligned}
    \label{delta_ceta_before}
\end{equation}

where:
\[
\alpha_1 = (2\pi)^{-(n+1)/2} \sigma_1^{-n} \sigma_2^{-1} 
\]

\[
A = \frac{n}{\sigma_1^2} + \frac{1}{\sigma_2^2}, \quad B = \frac{n\overline{x}}{\sigma_1^2} + \frac{\mu}{\sigma_2^2}, \quad C = \frac{\sum_{i=1}^{n} x_i^2}{\sigma_1^2} + \frac{\mu^2}{\sigma_2^2}.
\]


Simplifying, we get:
\begin{equation}
    \begin{aligned}
    p(x | \theta) \pi(\theta) = \alpha_2\exp\{ -\frac{A}{2}(\theta-\frac{B}{A})^2\}
\end{aligned}
    \label{delta_ceta}
\end{equation}

where \[
\alpha_2 = \alpha_1\exp\{\frac{B^2}{2A}-\frac{C}{2}\}
\]

The marginal probability \(p(x)\) is:

\begin{equation}
    \begin{aligned}
    p(x) &=\int p(x | \theta) \pi(\theta) d\theta = \alpha_2\sqrt{\frac{2\pi}{A}}
\end{aligned}
    \label{delta_ceta}
\end{equation}

\begin{algorithm}
\caption{Camel Algorithm}
\label{algorithm:TS}
\begin{algorithmic}[1]
\STATE{function MAIN()}{} \hspace{0.2cm} \footnotesize \textit{/* Main entry */} \normalsize
\FOR{$t = 1, 2, \dots, T$}
    \STATE $arm = \arg \min \mathrm{EVAL}()$
    \STATE Observe $cost$  by pulling $arm$
    \STATE $\mathrm{UPDATE}(cost, arm)$
\ENDFOR

{function EVAL()}
    \STATE \footnotesize \textit{/* SET represents the set of \(\theta\) for different arms.*/}\normalsize
    \STATE  $SET = \emptyset$ 
    \STATE \footnotesize \textit{/* n denotes the total number of arms. */}\normalsize
    \FOR{$i = 1, 2, \dots, n$}         
        \STATE  $\theta_i \sim \mathcal{N}(\mu_i, \sigma_{2,i}^2)$
        \STATE  $SET = SET \cup \{\theta_i\}$
    \ENDFOR
    \STATE \textbf{return} $SET$

{function UPDATE}(cost, arm) 
\STATE \footnotesize \textit{/* COST represents the set of costs for different arms.*/}\normalsize
    \STATE  $COST_{arm} = COST_{arm} \cup \{cost\}$
    \STATE  $\sigma_1 = \text{var}(COST_{arm})$
    \STATE  update $\mu, \sigma_2$ of $arm$ according to Eq. \ref{update_mu} and Eq. \ref{update_sigma}

\end{algorithmic}
\end{algorithm}

Thus, the posterior distribution is: 

\begin{equation}
    \begin{aligned}
    \pi (\theta | x) = &\frac{p(x | \theta) \pi(\theta)}{p(x)} = \sqrt{\frac{A}{2\pi}}\exp\{ -\frac{A}{2}(\theta-\frac{B}{A})^2\}
\end{aligned}
    \label{s}
\end{equation}

The posterior distribution is Gaussian \( \mathcal{N}(\widetilde{\mu}, \widetilde{\sigma}_2^2) \), where: 

\begin{equation}
    \begin{aligned}
   \widetilde{\mu} = \frac{B}{A} = \frac{\frac{n\overline{x}}{\sigma_1^2}+\frac{\mu}{\sigma_2^2}}{\frac{n}{\sigma_1^2}+\frac{1}{\sigma_2^2}} = \frac{n\xi_1\overline{x}+\mu\xi_2}{n\xi_1+\xi_2}
\end{aligned}
    \label{update_mu}
\end{equation}

\begin{equation}
    \begin{aligned}
   \widetilde{\sigma}_2^2 = \frac{1}{A} = \frac{1}{\frac{n}{\sigma_1^2}+\frac{1}{\sigma_2^2}} = \frac{1}{n\xi_1+\xi_2}
\end{aligned}
    \label{update_sigma}
\end{equation}
with \( \xi_1 = \frac{1}{\sigma_1^2} \) and \( \xi_2 = \frac{1}{\sigma_2^2} \).

By sampling \( x \) and using the prior distribution \( \pi(\theta) \), the posterior distribution \( \pi(\theta | x) \) can be iteratively updated. As the number of samples increases, the posterior distribution converges to the true distribution of \( \theta \).


\subsection{Algorithm of Camel}

How Camel works is detailed in Algorithm \ref{algorithm:TS}. The core of the Camel algorithm is the MAIN function,  which consists of two key functions: EVAL and UPDATE. The EVAL function samples from the \(\theta\) distribution of each arm to generate a set of \(\theta\) values. The UPDATE function selects the arm with the smallest \(\theta\) from this set and updates the posterior distribution based on the observed cost of executing that arm. The cost is assumed to follow a Gaussian distribution with a known variance \(\sigma_1\), which is estimated by collecting cost data for the corresponding arm. The detailed equation for updating the posterior distribution is provided in the prior section.


\section{Implementation and Evaluation}



\subsection{Implementation of Camel}
We implemented Camel on NVIDIA Jetson AGX Orin using Python 3.8 and the Llama.cpp framework. By varying the batch size (from 4 to 28 in increments of 4) and GPU frequency (7 categories), we generated 49 unique configurations. Each configuration corresponds to different inference latency and energy consumption, that is, various \texttt{cost}.
We set the \texttt{ngl} parameter in Llama.cpp to match the number of layers in the LLM, offloading all transformer layers to the GPU for computation. The \texttt{parallel} parameter was used to regulate the maximum request number for LLM batching, and we set it to 28. User requests were simulated using the \texttt{requests} library, and concurrent requests were handled using the \texttt{threadPoolExecutor} function. GPU frequency was adjusted by modifying the \texttt{/sys/class/devfreq/1700000.ga10b/max\_freq} and 
\texttt{/sys/class/devfreq/1700000.ga10b/min\_\\freq} files. Power consumption was monitored via the I2C interface at 100ms intervals.

\subsection{Evaluation Setup}


We conducted experiments on the Jetson AGX Orin platform using two models: Qwen2.5-3B \cite{qwen25} and Llama3.2-1B \cite{Llama32}, both quantized using the Q5\_K\_M method. Specifically, the operating system of the device is Ubuntu 20.04. The device is equipped with a CPU based on the Arm Cortex-A78AE architecture, a GPU from the NVIDIA Ampere series, and a memory capacity of 32GB. The default request arrival interval was one second. Within the Llama.cpp framework, maximum generated tokens were set to 70, and temperature was set to 0 for deterministic outputs. \textbf{The default \(\alpha\) in Eq. \ref{eq:objective} was 0.5}, while we also studied the impact of varying \(\alpha\).

The evaluation consisted of two parts: (1) Optimal configuration search: comparing Camel with grid search to find the best GPU frequency and batch size, evaluating the efficiency of each algorithm in identifying the optimal configuration. \textbf{The optimal configuration is the configuration with the minimum cost.} Both algorithms processed 3200 identical data points over 49 rounds; (2) Optimal configuration validation: testing Camel's optimal configuration against default configurations using the \texttt{alpaca} \cite{alpaca} dataset to accurately simulate real user request scenarios. 

The baselines included grid search for configuration search and three default configurations for optimal configuration validation: (max frequency, min batch size), (max frequency, max batch size), and (min frequency, max batch size).  (min frequency, min batch size) was not considered due to its poor performance.

We selected five evaluation metrics.
\textbf{Energy consumption} is the average energy consumed per request.
\textbf{Latency} is the average time from request arrival at the server to its final processing by the LLM.
\textbf{EDP} is the product of energy consumption and latency.
\textbf{Cost} combines power consumption and latency, as defined earlier. In each round of the optimal configuration search, cost is calculated from energy use and latency; a lower cost means a configuration is closer to optimal. The average cost over all rounds indicates the search quality. During configuration validation, cost is based on the average energy consumption and latency measured after each LLM batching.
\textbf{Regret} measures the difference between the current and the best possible result that could have been achieved by selecting the optimal configuration by comparing their cost values.


\subsection{Results1: Optimal Configuration Search}

\begin{figure}[htb]
\centering
    \includegraphics[width=0.9\linewidth]{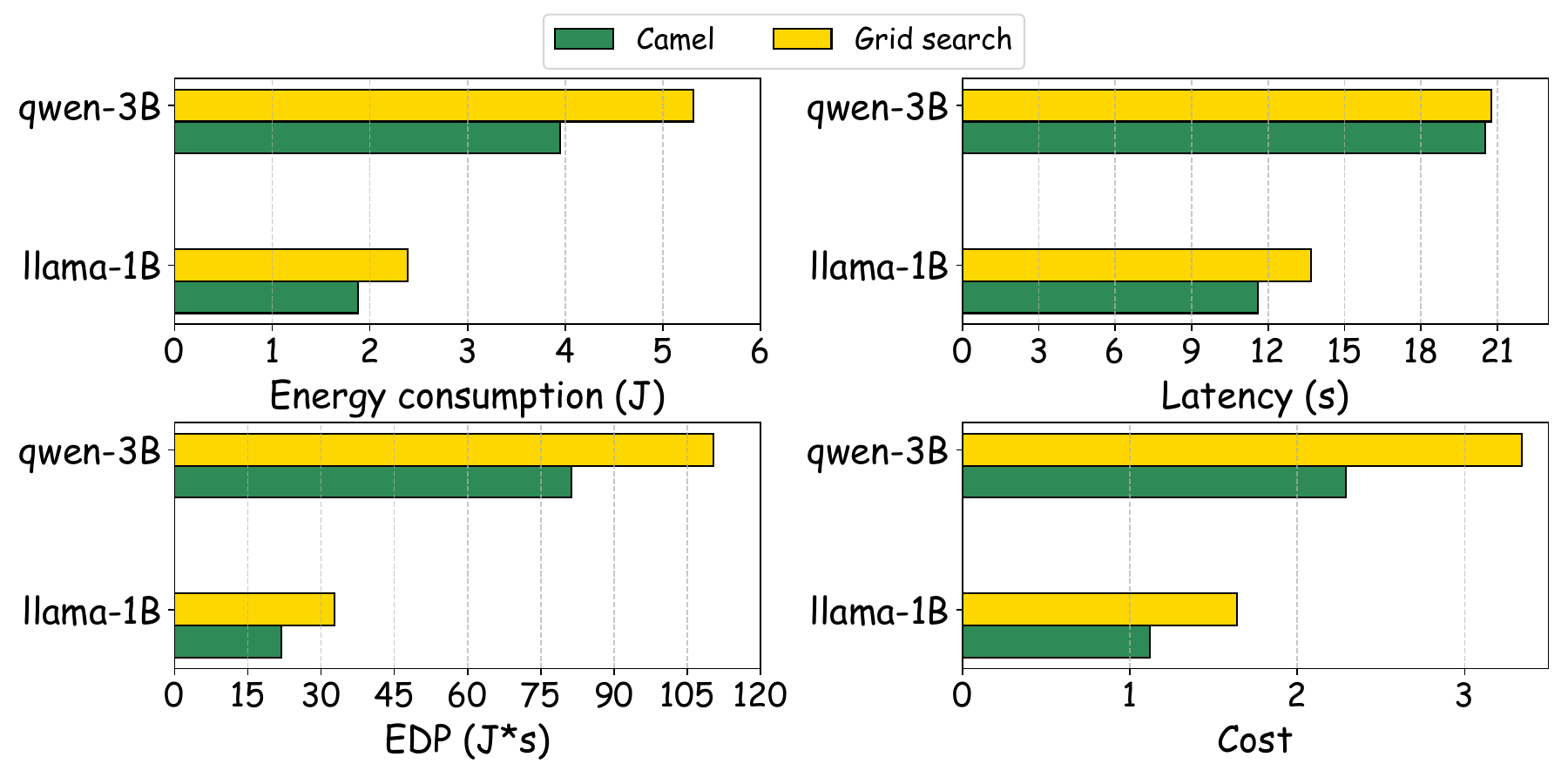}
    \caption{Performance of Camel and grid search in identifying the optimal configuration.}
    \label{fig:performance}
\end{figure}

Fig. \ref{fig:performance} compares performance metrics between grid search and Camel after 49 search rounds on two LLMs.

\begin{figure}[htb]
\centering
\includegraphics[width=1\linewidth]{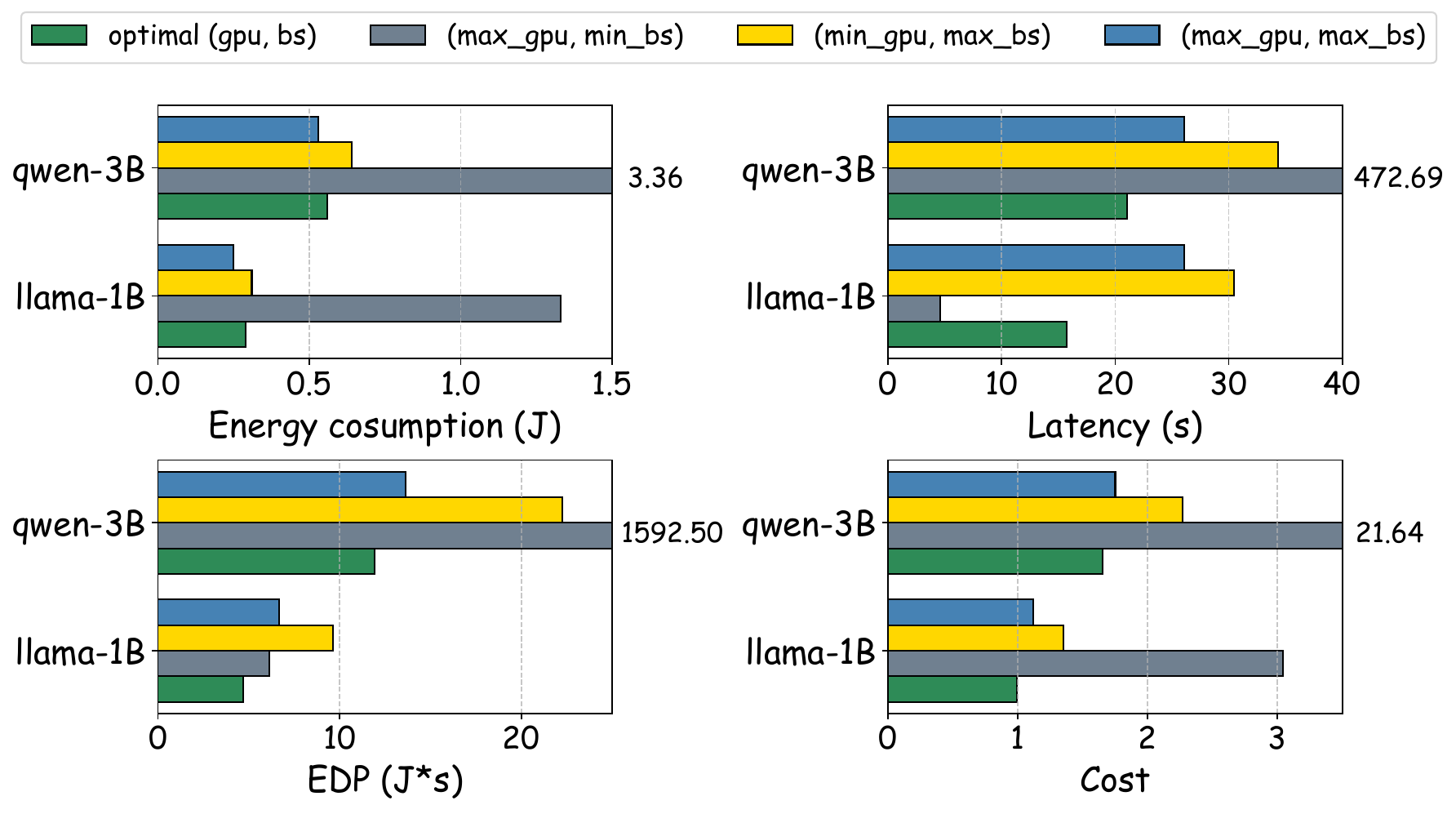}
    \caption{Comparison of performance between the optimal and three default configurations for LLM inference.}
    \label{fig:optimal}
\end{figure}

\begin{figure}[htb]
    \centering
    \subfigure[Cumulative regret (LLaMA)]{%
        \includegraphics[width=0.49\linewidth]{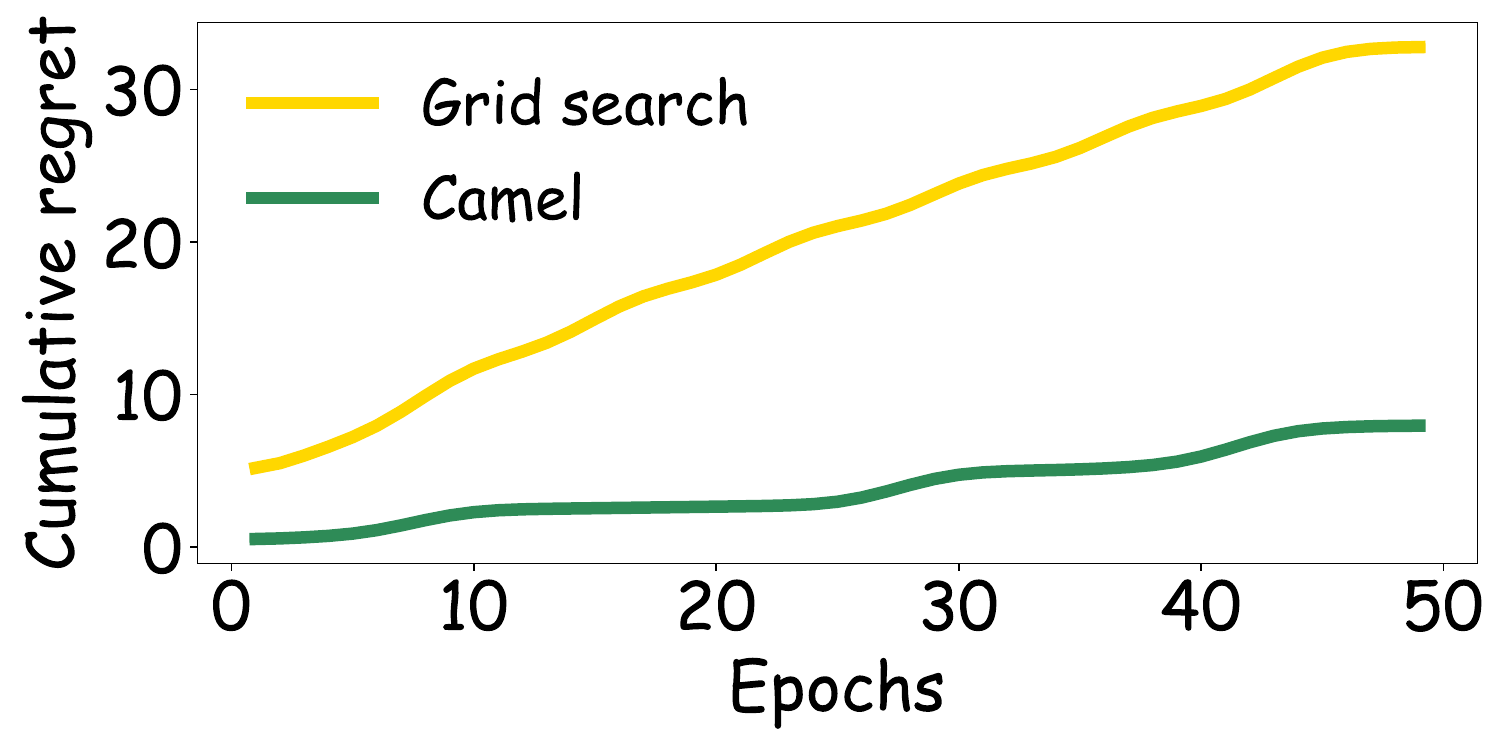}%
        \label{fig:llama-regret}%
    }\hspace{0pt}
    \subfigure[Cumulative regret (Qwen)]{%
        \includegraphics[width=0.49\linewidth]{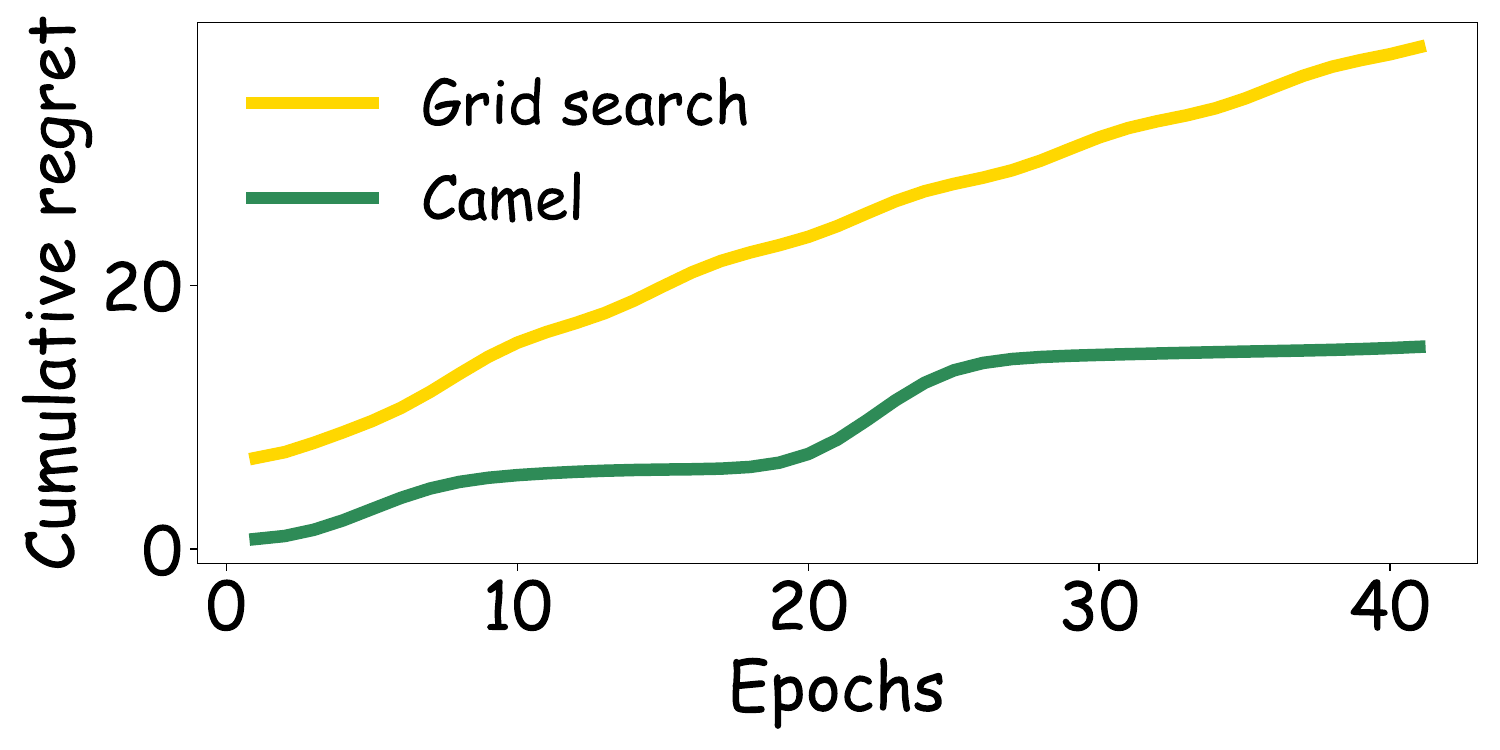}%
        \label{fig:qwen-regret}%
        }
    \caption{Cumulative regret of grid search and Camel in finding the optimal configuration.}
    
    \label{fig:regret}
\end{figure}

\begin{figure}[htb]
    \centering
    \subfigure[The complete search process on LLaMA]{%
        \includegraphics[width=1\linewidth]{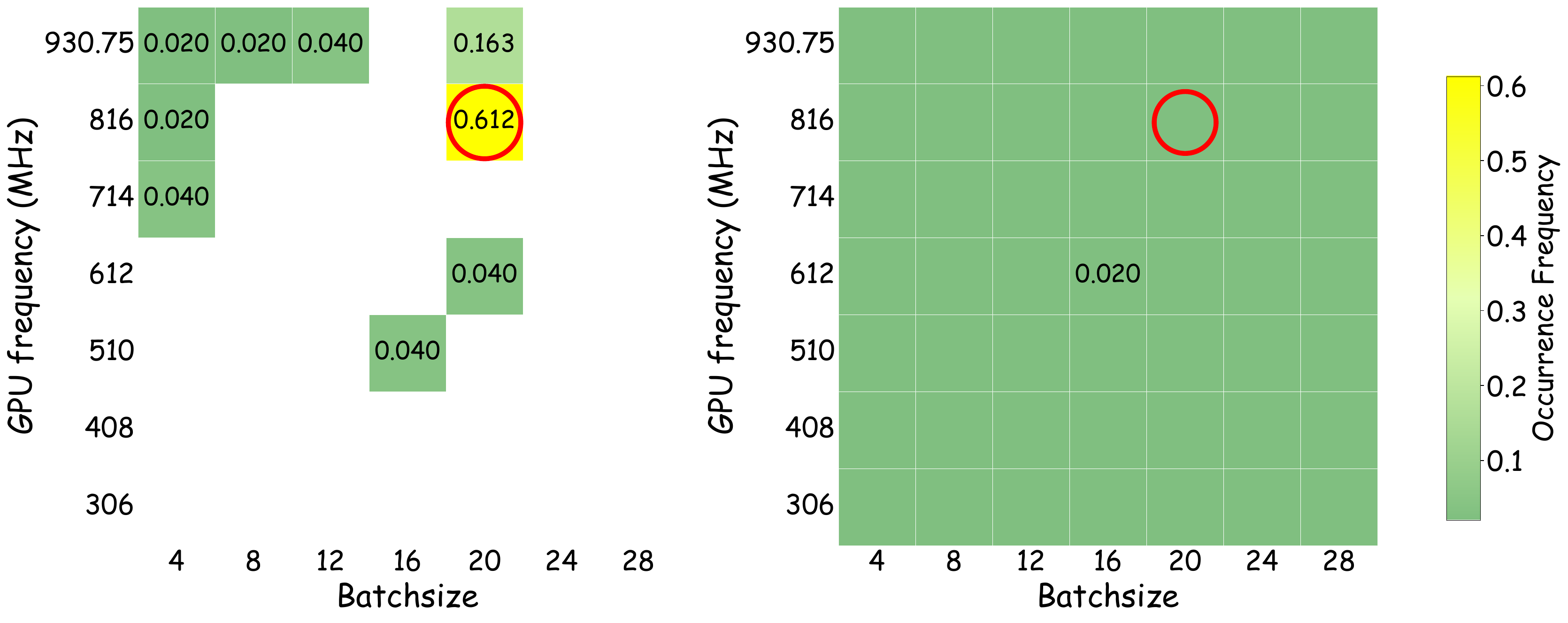}%
        \label{fig:heatmap1}%
    }\hspace{0pt}
    \subfigure[The complete search process on Qwen]{%
        \includegraphics[width=1\linewidth]{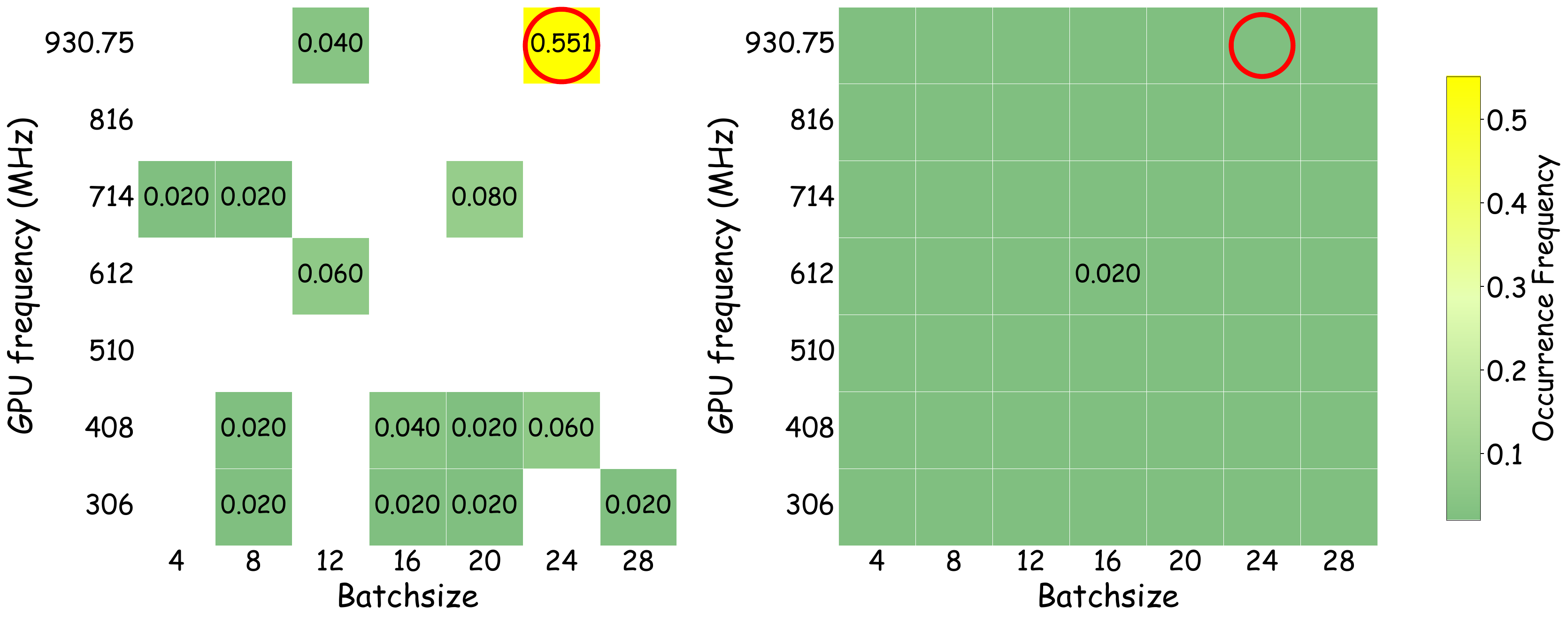}%
        \label{fig:heatmap2}%
        }
    \caption{The frequency of configurations explored by Camel and grid search during the search for the optimal configuration on LLaMA (a) and Qwen (b), with the red circle indicating the optimal configuration.}
\end{figure}



\textbf{Energy consumption.}
Camel reduces energy consumption by 27.13\% (Llama3.2-1B) and 34.43\% (Qwen2.5-3B) compared to grid search. Notably, as the model size increases (Qwen2.5-3B), energy consumption rises for both algorithms, reflecting greater computational demands.

\textbf{Latency.}
Camel reduces latency by 17.93\% (Llama3.2-1B) and 1.12\% (Qwen2.5-3B) compared to grid search. Similar to energy consumption, latency increases for both algorithms on the Qwen2.5-3B model.

\textbf{EDP.} Camel reduces EDP by 49.45\%  (Llama3.2-1B) and 35.75\% (Qwen2.5-3B) compared to grid search. When considering both energy consumption and inference latency, Camel demonstrates a more efficient search process than grid search.

\textbf{Cost.}
Camel reduces cost by 46.43\%  (Llama3.2-1B) and 45.85\% (Qwen2.5-3B) compared to grid search. This indicates that, compared to grid search, Camel incurs a lower cost at each step during the process of identifying the optimal configuration.

\textbf{Cumulative regret.}
Fig. \ref{fig:regret} shows cumulative regret of grid search is 3.8x (Llama3.2-1B) and 2.3x (Qwen2.5-3B) higher than Camel. Indicating Camel finds the optimal configuration faster and at lower cost.

\textbf{Selections during search.}
Fig. \ref{fig:heatmap1}  and Fig. \ref{fig:heatmap2} shows exploration frequencies by both algorithms across 49 search rounds. Grid search explores uniformly (0.02 frequency per configuration). Both algorithms converge on (930.75MHz, 24) for Qwen2.5-3B and (816MHz, 20) for Llama3.2-1B. Camel explored more configurations on the Qwen2.5-3B model, resulting in higher regret (as shown in Fig. \ref{fig:qwen-regret}).

\begin{figure*}[htbp]
	\centering
	\begin{minipage}[t]{0.24\linewidth}
		\centering
		\includegraphics[width=1\linewidth]{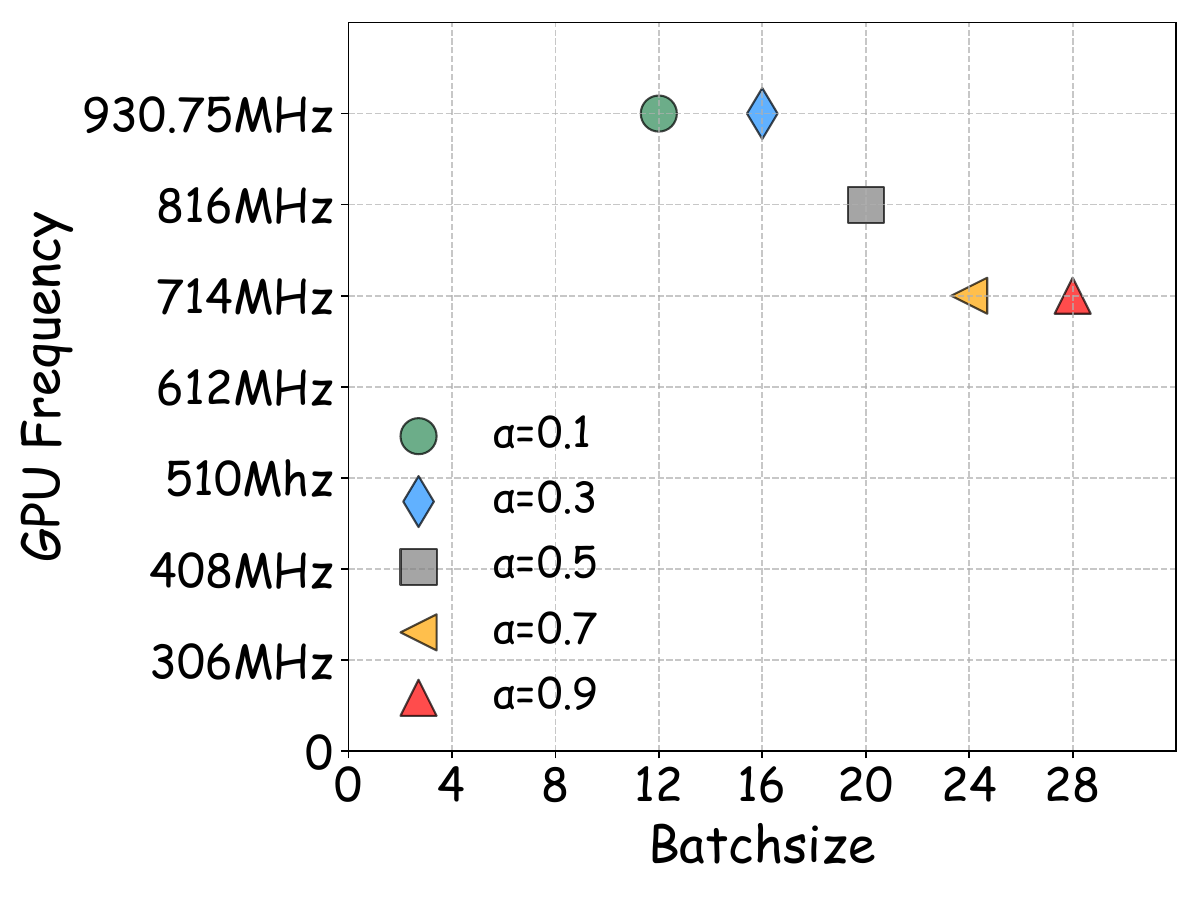}
		\caption{The optimal GPU frequency and batch size under different $\alpha$ values.}
		\label{fig:alpha}
	\end{minipage}
	\begin{minipage}[t]{0.24\linewidth}
		\centering
	\includegraphics[width=1\linewidth]{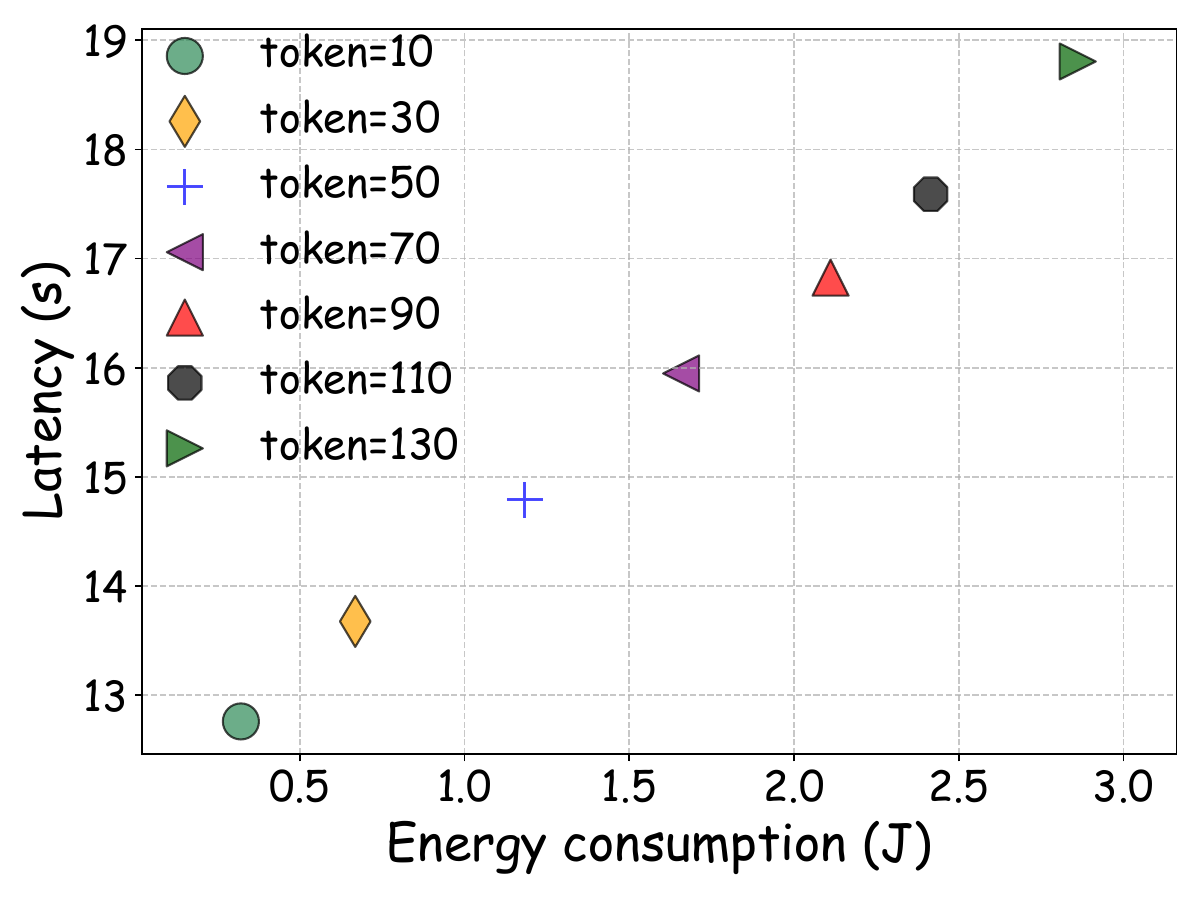}
		\caption{Energy consumption and latency for varying maximum generated tokens with (max\_{GPU}, max\_{bs}).}
		\label{fig:token}
	\end{minipage}
        \begin{minipage}[t]{0.24\linewidth}
		\centering
		\includegraphics[width=1\linewidth]{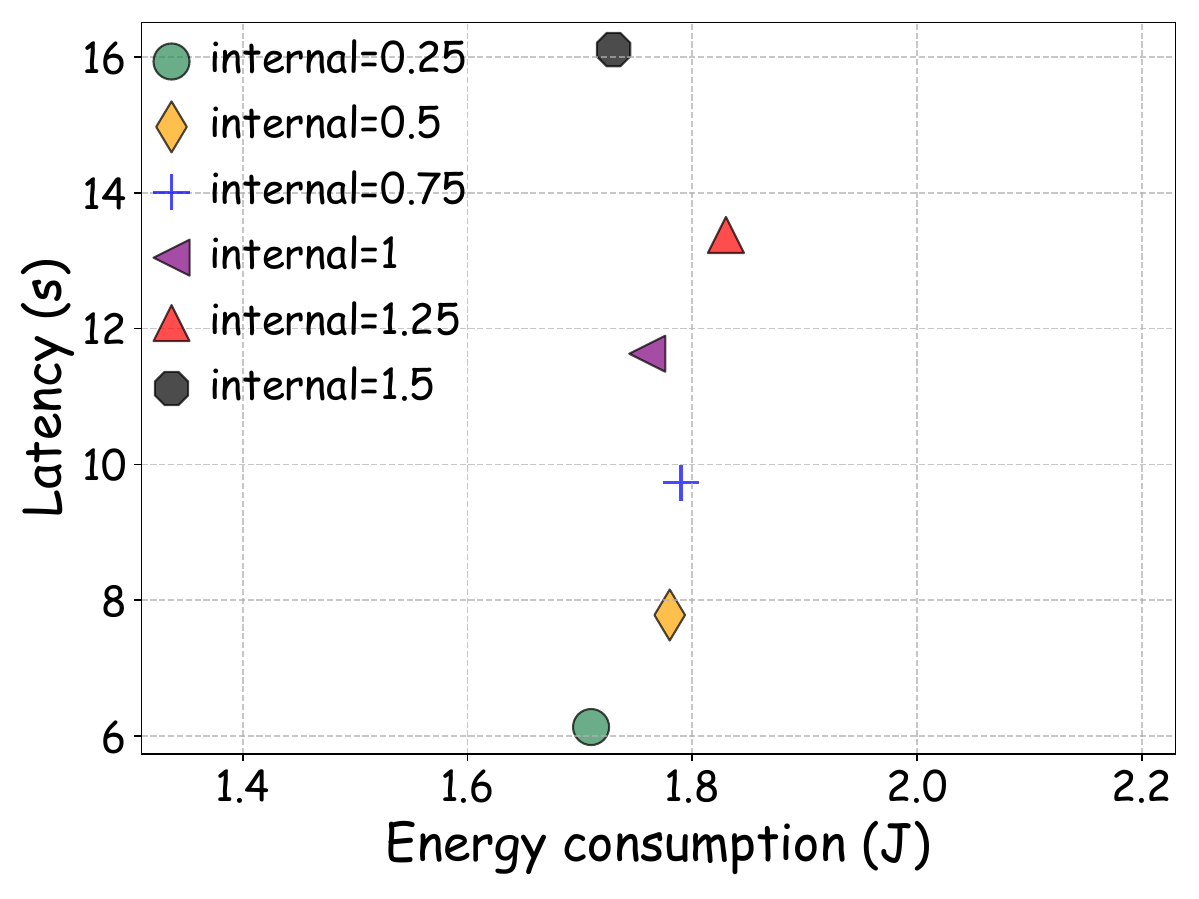}
		\caption{Energy consumption and latency of the optimal configuration under different internal times.}
		\label{fig:internal}
	\end{minipage}
        \begin{minipage}[t]{0.24\linewidth}
		\centering
		\includegraphics[width=1\linewidth]{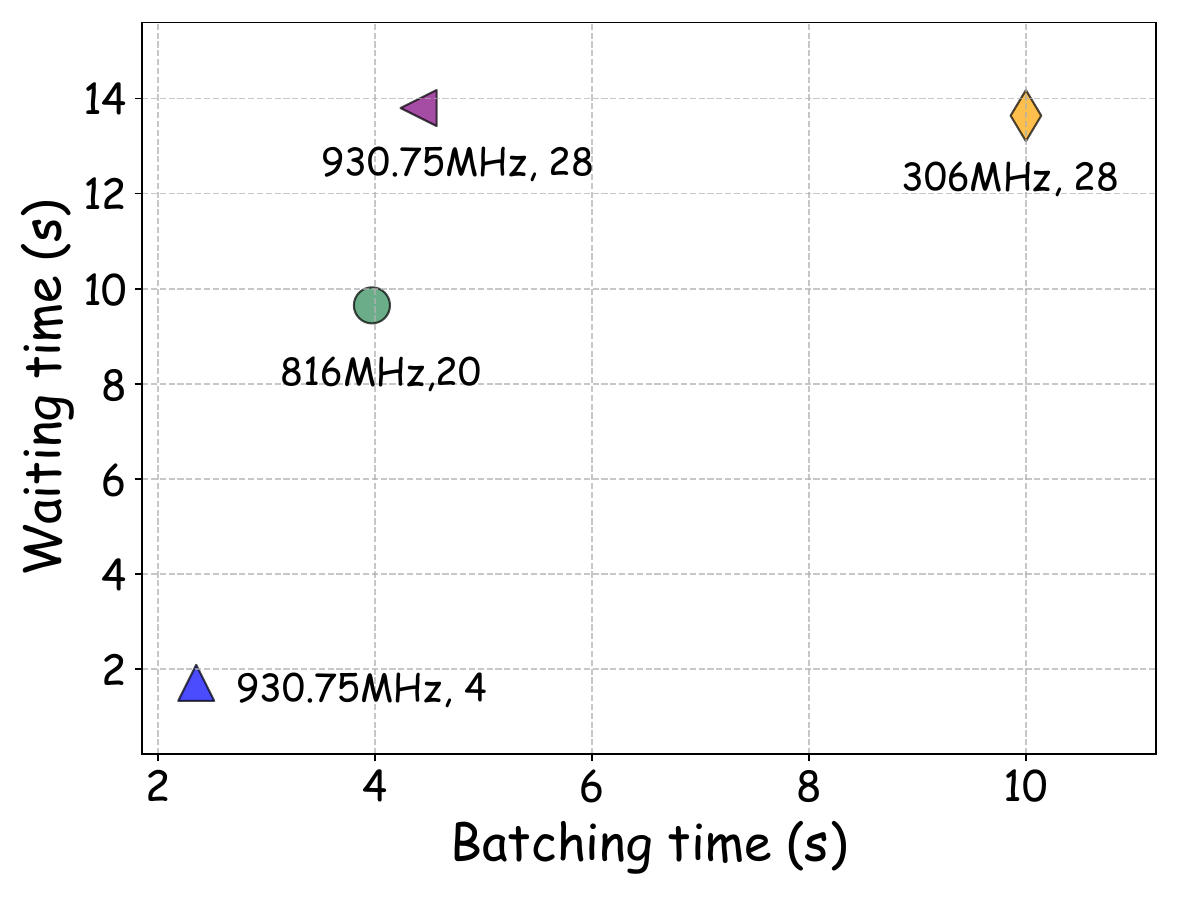}
		\caption{Batching time and waiting time under various configurations on LLaMA.}
		\label{fig:waiting}
	\end{minipage}
\end{figure*}

\subsection{Results2: Optimal Configuration Validation}

\subsubsection{Comparison with Default Configurations}

We compared the optimal configuration identified by Camel with three default configurations using 2500 samples from the \texttt{alpaca} dataset. The results are shown in Fig. \ref{fig:optimal}.

\textbf{Energy consumption.} The optimal configuration of Camel reduced energy consumption to 21.8\% (Llama3.2-1B) and 16.67\% (Qwen2.5-3B) of (max frequency, min batch size). Compared to (min frequency, max batch size), energy consumption decreased by 6.45\% and 12.5\%, respectively. However, it increased by 16.67\% and 5.66\% compared to (max frequency, max batch size), indicating that the optimal configuration of Camel is not solely focused on minimizing energy consumption but rather balances both energy consumption and latency, as further demonstrated in the EDP and cost analyses.

\textbf{Latency.} Latency decreased by 48.24\% (Llama3.2-1B) and 38.81\% (Qwen2.5-3B) compared to (min frequency, max batch size), and by 39.56\% and 17.99\% compared to (max frequency, max batch size). On Qwen2.5-3B, the optimal configuration significantly reduced latency compared to (max frequency, min batch size). However, on Llama3.2-1B, the latency increased to 3.4x that of (max frequency and min batch size). This is because Qwen2.5-3B, being a larger model, has longer batch processing times (5.49s). When requests are sent at 1-second intervals, a batch-sized request causes a 4-second delay (as the minimum batch size is 4). This, combined with the longer batch processing time (5.49s), creates a ``bottleneck", increasing waiting times for subsequent requests. In contrast, Llama3.2-1B, a smaller model with lighter computational load and shorter batch processing times (2.86s), avoids this ``bottleneck" since the 4-second interval between batch requests is sufficient.

\textbf{EDP.} The optimal configuration of Camel achieved the minimum EDP for both models. Compared to (min frequency, max batch size), it reduced EDP by 51.35\% and 46.34\%, respectively. Against (max frequency, max batch size), EDP was reduced by 29.94\% and 12.46\%. For Llama3.2-1B, EDP was 23.9\% lower than (max frequency, min batch size), while Qwen2.5-3B saw a more significant EDP reduction.
The optimal configuration of Camel achieved the best overall performance by considering both energy consumption and latency. 

\textbf{Cost.} The optimal configuration achieved the lowest average cost for both Llama3.2-1B and Qwen2.5-3B. Compared to (max frequency, min batch size), costs were reduced to 32.57\% and 7.66\%, respectively. Against (min frequency, max batch size), costs decreased by 26.67\% and 27.08\%. Lastly, compared to (max frequency, max batch size), it reduced by 11.61\% and 5.41\%, respectively.

\subsubsection{In-depth Analysis on Latency}

Based on the analysis from Eq. \ref{eq:total_latency}, latency comprises two components: queue waiting time for requests and batching time. Fig. \ref{fig:waiting} presents the average batching time and waiting time for each request across four configurations on Llama3.2-1B. (930.75MHz, 28) represents the highest GPU frequency and batch size, while (306MHz, 28) represents the lowest frequency and the highest batch size. (930.75MHz, 4) uses the highest frequency and smallest batch, and (816MHz, 20) is the optimal configuration.
Increasing GPU frequency from 306MHz to 930.75MHz (at batch size 28) reduces batching time by 56\%, without affecting waiting time. Reducing batch size from 28 to 4 (with frequency at 930.75MHz) decreases batching time by 46.5\% and significantly reduces waiting time, matching the theoretical analysis in Fig. \ref{eq:total_latency}. Finally, comparing (930.75MHz, 28) and (816MHz, 20), although the GPU frequency decreases, reducing batch size still results in a 9.7\% reduction in batching time, demonstrating how batch size and GPU frequency jointly influence latency.

\subsubsection{Sensitivity of \(\alpha\)}
Fig. \ref{fig:alpha} shows how the optimal batch size and GPU frequency change with varying \(\alpha\) values during Llama3.2-1B inference. As \(\alpha\) increases, the weight of energy consumption optimization in the objective function (Eq. \ref{eq:objective}) increases, while the weight of latency optimization decreases. As shown in Fig. \ref{fig:alpha}, increasing \(\alpha\) leads to a decrease in GPU frequency and an increase in batch size.  This indicates that a larger batch size and lower GPU frequency reduce energy consumption, while higher GPU frequencies and smaller batch sizes are better for minimizing latency.

\subsubsection{Sensitivity of Internal Time}
Fig. \ref{fig:internal} shows how average energy consumption and latency change during Llama3.2-1B inference with optimal GPU frequency and batch size, at varying request arrival intervals. As the interval increases, energy consumption stabilizes, while latency rises. According to Eq. \ref{eq:total_objective}, the increase in arrival intervals results in a decrease in the request arrival rate, $\lambda$, which leads to longer waiting times and an increase in $L_{request}$.

\subsubsection{Sensitivity of Token Length}
Fig. \ref{fig:token} shows how average energy consumption and latency change during Llama3.2-1B inference with maximum GPU frequency and batch size, as token generation length varies. As the number of tokens increases, both energy consumption and latency increase linearly. This aligns with the analysis in Fig. \ref{eq:total_objective}, where, increasing \(c_p\) leads to  higher $E_{request}$ and $L_{request}$.



\section{Conclusion}
Energy consumption is a key factor when performing LLM inference on edge devices. In addition to energy consumption, inference latency must also be carefully considered. Thus, finding the right balance between energy use and latency through proper configuration is essential. To address this challenge, this paper proposes Camel, an energy consumption management framework. By effectively tackling the exploration-exploitation dilemma, the framework can rapidly identify the optimal GPU frequency and batch size during LLM batching. Experimental results show that compared to the default configuration, the optimal configuration identified by Camel significantly reduces EDP, achieving a better balance between energy consumption and latency.

\bibliography{aaai2026}

\end{document}